\documentclass[a4paper,11pt]{article}

\usepackage[T1]{fontenc}
\usepackage[english]{babel}

\usepackage[margin=25mm]{geometry}

\usepackage{amsmath,amssymb}

\newcommand*{\bs}{\boldsymbol}
\newcommand*{\mcal}{\mathcal}
\newcommand*{\mbb}{\mathbb}

\newcommand*{\diff}{\mathop{}\!\mathrm{d}}
\newcommand*{\pd}{\partial}
\newcommand*{\e}{\mathop{\mathrm{e}}\nolimits}
\newcommand*{\const}{\ensuremath{\mathrm{const}}}
\DeclareMathOperator{\gradr}{\bs\nabla\!}
\DeclareMathOperator{\diverg}{div\!}
\newcommand*{\defin}{\stackrel{\mathrm{def}}{=}}
\newcommand*{\Id}{\mathcal{I}}
\newcommand*{\total}{{}}
\newcommand*{\scatt}{\mathrm{s}}
\newcommand*{\average}[1]{\big<#1\big>}
\newcommand*{\Lapl}{\mathcal{L}}
\newcommand*{\spaceR}{\mathbb{R}^3}
\newcommand*{\spaceV}{\mathbb{V}}
\newcommand*{\sphere}{{\mathbb{S}^2}}
\newcommand*{\outersol}{\mathrm{o}}
\newcommand*{\innersol}{\mathrm{i}}

\usepackage{xcolor}
\usepackage{indentfirst}
\usepackage[perpage, symbol]{footmisc}

\usepackage[numbers, sort&compress]{natbib}
\newcommand*{\ie}{i.\,e.}

\numberwithin{equation}{section}
\allowdisplaybreaks[4]

\usepackage[unicode,colorlinks]{hyperref}

\definecolor{darkblue}{rgb}{0,0,0.7}

\hypersetup{%
  bookmarksopen=true,
  bookmarksopenlevel=1,
  bookmarksnumbered=true,
  linkcolor=darkblue,
  citecolor=darkblue,
  pdfpagelayout=OneColumn,
  pdfstartview={XYZ null null 1},
  pdfview={XYZ null null null}
}

\begin{document}

\title{%
  \bfseries\LARGE
  Linear Boltzmann-like equation, describing non-classical particle transport, and related asymptotic solutions for small mean free paths
}

\author{%
  \large
  Sergey~A.~Rukolaine\,$^{1,2,}$\thanks{E-mail address: \texttt{rukol@ammp.ioffe.ru}}
}

\date{%
  \itshape\small
  $^1$\,Ioffe Institute, 26 Polytekhnicheskaya, St.\,Petersburg 194021, Russia\\
  $^2$\,St.\,Petersburg State Polytechnical University, 29 Polytekhnicheskaya, St.\,Petersburg 195251, Russia
}

\maketitle

\begin{abstract}
  In classical kinetic or kinetic-like models a particle free path distribution is exponensial, but this is more likely to be an exception than a rule. In this paper we derive a linear Boltzmann-like equation for a general free path distribution in the framework of Alt's model J. Math. Biol. \textbf{9}, 147 (1980). In the special case that the free path distribution has at least first and second finite moments we construct an asymptotic solution of the equation for small mean free paths. The asymptotic solution becomes a diffusion approximation to the one-speed Boltzmann-like equation.
\end{abstract}

\section{Introduction}

Classical neutron transport or transport of thermal energy by photons is described by the linear Boltzmann kinetic equation \cite{DuderstadtMartin:1979, Cercignani:1988, Modest:2013}, where the free path distribution is exponential, \ie, collision events form the Poisson process \cite{Ross:2007}.
This is a very restrictive assumption, and, therefore, in many cases it is violated \cite{Uchaikin:1998, DavisMineevWeinstein:2011, LarsenVasques:2011, Uchaikin:2013}.
Kinetic and kinetic-like models are also used for description of transport processes in biology \cite{OthmerEtAl:1988, HillenOthmer:2000, OthmerHillen:2002, BellouquidDelitala:2006, Bellomo:2008, BiancaBellomo:2011, BellomoEtAl:2012, BressloffNewby:2013, Bressloff:2014}. The common assumption in these models is also the exponential distribution of the free path, but this is more likely to be an exception than a rule.

A model, which extends the linear Boltzmann equation to a general free path distribution, was proposed in Ref.~\cite{Alt:1980}. This model was embedded in a biological context. The model does not take into account absorption and sources, but their inclusion is straightforward.
In Ref.~\cite{Alt:1980} a partial integro-differential kinetic equation and its diffusion approximation (Patlak-Keller-Segel equation) were derived under the assumption that the free path distribution was asymptotically exponential.
The steady-state case of Alt's model~\cite{Alt:1980} in a physical context was rederived in Ref.~\cite{LarsenVasques:2011}.
A more general model, including acceleration of the particles due to external forces, was proposed in Ref.~\cite{FriedrichEtAl:2006_PRL}, see also Ref.~\cite{FriedrichEtAl:2006_PRE}, where a Boltzmann-like equation without absorption and sources was obtained.

Alt's model~\cite{Alt:1980} can be considered in the framework of the velocity jump model of the continuous-time random walks (CTRWs)~\cite{OthmerEtAl:1988, ZumofenKlafter:1993}.
A model similar to Alt's one in the framework of the position jump model of the CTRWs was proposed in Ref.~\cite{VladRoss:2002}, see also Ref.~\cite{MendezEtAl:2010}.

In this paper we derive a linear Boltzmann-like equation for a general free path distribution in the framework of Alt's model. In the special case that the free path distribution has at least first and second finite moments we construct an asymptotic solution of the equation for small mean free paths like in Ref.\,\cite{LarsenKeller:1974}, where an asymptotic solution was constructed for the ordinary linear Boltzmann equation.
The asymptotic solution becomes a diffusion approximation to the one-speed Boltzmann-like equation. In a steady state this diffusion approximation coincides with that obtained in Ref.~\cite{FrankGoudon:2010} and differs from the diffusion approximation in Ref.~\cite{LarsenVasques:2011} by small terms of the second order.

All the reasoning in this paper is performed in the three-dimensional space. A generalization to the $n$-dimensional space is straightforward.

The paper is organized as follows.
Section~\ref{sec:Model} describes the model of non-classical particle transport.
In Section~\ref{sec:BoltzLikeEq} we derive the Boltzmann-like equation, which describes transport with an arbitrary distribution of the free path.
In Section~\ref{sec:AsySolution} we construct an asymptotic solution of the equation for small mean free paths.
In Section~\ref{sec:DiffApprox} the diffusion approximation to the one-speed Boltzmann-like equation, as a particular case of the asymptotic solution, is obtained.

\section{Non-classical particle transport: a model}
\label{sec:Model}

The model generalizes that of classical linear transport \cite{DuderstadtMartin:1979, Modest:2013}, where the extinction and scattering coefficients (total and scattering cross sections) do not depend on a free path travelled by a particle. In the non-classical transport the coefficients depend on the free run path. In this case the phase space density of particles takes the form $\xi(\bs{r}, \bs{v}, t, l)$, where $\bs{r}$ is a position, $\bs{r} \in \spaceR$, $\bs{v}$ is a velocity vector, $\bs{v} \in \spaceV \subset \spaceR$, $t \geq 0$ is time, $l \geq 0$ is the free path of the particle.
The phase space density obeys the equation~\cite{Alt:1980}
\begin{equation}
  \label{eq:PDEDensA}
  \pd_t^{} \xi
  + v \pd_l^{} \xi
  + \bs{v} \cdot \gradr\xi
  + v \sigma_\total^{} \xi
  = 0,
\end{equation}
where $\gradr$ means the gradient with respect to $\bs{r}$, $\sigma_\total^{} \equiv \sigma_\total^{}(\bs{r}, \bs{v}, l)$ is the extinction coefficient (or total cross section), $v = |\bs{v}|$.
This equation describes noninteracting particles moving at the point $\bs{r}$ with the velocity $\bs{v}$ so that they stop their free run (scatter or disappear) with the rate $v \sigma_\total^{}$.
We do not consider here particle acceleration due to external forces; taking it into account is straightforward.

Eq.\,\eqref{eq:PDEDensA} is supplemented by the condition, describing the density of the particles at the beginning of their free run,
\begin{equation}
  \label{eq:BoundCondDensA}
  \left. \xi \right|_{l=0}^{}
  \equiv
  \eta(\bs{r}, \bs{v}, t)
  =
  \mcal{K} \left[ \int_0^\infty \sigma_\scatt^{}(\bs{r}, \bs{v}, l') \,\xi(\bs{r}, \bs{v}, t, l') \diff l' \right]
  + F(\bs{r}, \bs{v}, t)
\end{equation}
where
\begin{equation}
  \label{eq:ScattOperator}
  \mcal{K} f(\bs{r}, \bs{v}, t)
  \defin
  \int_{\spaceV}^{} K(\bs{r}, \bs{v}, \bs{v}') f(\bs{r}, \bs{v}', t) \diff\bs{v}'
\end{equation}
is the scattering operator,
$K(\bs{r}, \bs{v}, \bs{v}') \equiv K(\bs{r}, \bs{v}' \to \bs{v})$ is the scattering kernel such that
\begin{equation*}
  K \geq 0
  \quad\text{and}\quad
  \int_{\spaceV}^{} K(\bs{r}, \bs{v}, \bs{v}') \diff\bs{v} = 1
\end{equation*}
[\ie, $K(\bs{r}, \cdot, \bs{v}')$ is a probability density function],
$\sigma_\scatt^{} \equiv \sigma_\scatt^{}(\bs{r}, \bs{v}, l)$ is the scattering coefficient (or scattering cross section), $\sigma_\scatt^{} \leq \sigma_\total^{}$,
$F$ is the source term.
The condition~\eqref{eq:BoundCondDensA} is a straightforward modification of the corresponding condition given in Ref.~\cite{Alt:1980}.
This condition means that the particles at the beginning of their free run arise as a result of scattering and are generated by sources. This condition means also that, if $\sigma_\scatt^{} < \sigma_\total^{}$, a part of the particles disappears (is absorbed).

In general case the phase space density is subject to the initial condition
\begin{equation*}
  \left. \xi \right|_{t=0}^{}
  =
  \vartheta(\bs{r}, \bs{v}, l),
\end{equation*}
where $\vartheta$ is an initial distribution of the density.
However, here we impose the initial condition
\begin{equation}
  \label{eq:InitCondDensADelta}
  \left. \xi \right|_{t=0}^{}
  =
  \chi(\bs{r}, \bs{v}) \delta(l),
\end{equation}
where $\delta(\cdot)$ is the Dirac delta function. This condition means that at the moment $t=0$ all the particles begin their free run.

\section{The Boltzmann-like equation}
\label{sec:BoltzLikeEq}

We define the phase space density
\begin{equation}
  \label{eq:DensT}
  \psi(\bs{r}, \bs{v}, t)
  =
  \int_0^\infty  \xi(\bs{r}, \bs{v}, t, l) \diff l,
\end{equation}
which does not take into account the free path.
Eqs.\,\eqref{eq:PDEDensA} and \eqref{eq:BoundCondDensA} yield
\begin{multline}
  \label{eq:PDEDensity}
  \pd_t^{} \psi
  + \bs{v} \cdot \gradr\psi\\
  =
  \mcal{K} \left[ v \int_0^\infty \sigma_\scatt^{}(\bs{r}, \bs{v}, l) \,\xi(\bs{r}, \bs{v}, t, l) \diff l \right]
  - v \int_0^\infty \sigma_\total^{}(\bs{r}, \bs{v}, l) \,\xi(\bs{r}, \bs{v}, t, l) \diff l
  + F(\bs{r}, \bs{v}, t)
\end{multline}
(we assume that $\xi \to 0$ as $l \to \infty$).

To express the right-hand side of Eq.\,\eqref{eq:PDEDensity} through $\psi$ we integrate Eq.\,\eqref{eq:PDEDensA} along the characteristics.
This yields
\begin{equation}
  \label{eq:SolutionOfPDE}
  \xi(\bs{r}, \bs{v}, t, l)
  =
  \begin{cases}
  S(\bs{r}, \bs{v}, l) \,\eta \!\left( \bs{r} - \bs\varOmega l, \bs{v}, t - \dfrac{l}{v} \right), &
  l < v t,\\[1ex]
  S(\bs{r}, \bs{v}, l) \,\chi(\bs{r} - \bs\varOmega l, \bs{v}) \delta(l - v t), &
  v t \leq l,
  \end{cases}
\end{equation}
where
\begin{equation}
  \label{eq:SurvivalProb}
  S(\bs{r}, \bs{v}, l)
  =
  \exp \left\{ - \int_0^l \sigma_\total^{}(\bs{r} - \bs\varOmega (l - l'), \bs{v}, l') \diff l' \right\}
\end{equation}
is the survival probability (see Appendix~\ref{sec:DensityOnCharacteristic} for details).
Therefore,
\begin{multline}
  \label{eq:IntSigmaDensA}
  \int_0^\infty \sigma_\total^{}(\bs{r}, \bs{v}, l) \,\xi(\bs{r}, \bs{v}, t, l) \diff l\\
  =
  \int_0^{v t} p(\bs{r}, \bs{v}, l) \,\eta \!\left( \bs{r} - \bs\varOmega l, \bs{v}, t - \dfrac{l}{v} \right) \diff l
  +
  p(\bs{r}, \bs{v}, v t) \chi(\bs{r} - \bs{v} t, \bs{v}),
\end{multline}
where
\begin{equation}
  \label{eq:PDF}
  p(\bs{r}, \bs{v}, l)
  =
  \sigma_\total^{}(\bs{r}, \bs{v}, l) S(\bs{r}, \bs{v}, l)
\end{equation}
is the probability density function (PDF) of a free path.
Eq.\,\eqref{eq:IntSigmaDensA}, the Laplace transform and the equality $h(\bs{r} - \bs\varOmega l) = \e^{-l \bs\varOmega \cdot \gradr} h(\bs{r})$ yield
\begin{multline*}
  \Lapl \!\left[ \int_0^\infty \sigma_\total^{}(\bs{r}, \bs{v}, l) \,\xi(\bs{r}, \bs{v}, t, l) \diff l \right]\\
  \equiv
  \Lapl \!\left[ \int_0^{v t} p(\bs{r}, \bs{v}, l) \e^{-l \bs\varOmega \cdot \gradr} \eta \!\left( \bs{r}, \bs{v}, t - \dfrac{l}{v} \right) \diff l
  + p(\bs{r}, \bs{v}, v t) \e^{-v t \bs\varOmega \cdot \gradr} \chi(\bs{r}, \bs{v}) \right]\\
  =
  \frac{1}{v} \Lapl{p} \!\left( \bs{r}, \bs{v}, \frac{s}{v} + \bs\varOmega \cdot \gradr \right) \left[ \Lapl\eta(\bs{r}, \bs{v}, s) + \chi(\bs{r}, \bs{v}) \right],
\end{multline*}
where
\begin{equation*}
  \Lapl{f}(s)
  =
  \int_0^\infty \e^{-s t} f(t) \diff t
  \quad\text{or}\quad
  \Lapl{g}(\lambda)
  =
  \int_0^\infty \e^{-\lambda l} g(l) \diff l
\end{equation*}
is the Laplace transform (we distinguish functions of $t$ and $l$).

Eqs.\,\eqref{eq:DensT} and \eqref{eq:SolutionOfPDE} yield
\begin{equation*}
  \psi(\bs{r}, \bs{v}, t)
  =
  \int_0^{v t} S(\bs{r}, \bs{v}, l) \,\eta \!\left( \bs{r} - \bs\varOmega l, \bs{v}, t - \dfrac{l}{v} \right) \diff l
  +
  S(\bs{r}, \bs{v}, v t) \chi(\bs{r} - \bs{v} t, \bs{v}),
\end{equation*}
or, equivalently,
\begin{equation*}
  \Lapl\psi(\bs{r}, \bs{v}, s)
  =
  \frac{1}{v} \Lapl{S} \!\left( \bs{r}, \bs{v}, \frac{s}{v} + \bs\varOmega \cdot \gradr \right) \left[ \Lapl\eta(\bs{r}, \bs{v}, s) + \chi(\bs{r}, \bs{v}) \right].
\end{equation*}
Finally, if we define the memory kernel $\phi(\bs{r}, \bs{v}, l)$ by~\cite{KenkreEtAl:1973}
\begin{equation}
  \label{eq:MemoryKernel}
  \Lapl\phi_\total^{}(\bs{r}, \bs{v}, \lambda)
  \defin
  \frac{\Lapl{p}(\bs{r}, \bs{v}, \lambda)}{\Lapl{S}(\bs{r}, \bs{v}, \lambda)}
  \equiv
  \frac{\lambda \Lapl{p}(\bs{r}, \bs{v}, \lambda)}{1 - \Lapl{p}(\bs{r}, \bs{v}, \lambda)},
\end{equation}
which is equivalent to
\begin{equation*}
  p(\bs{r}, \bs{v}, l)
  =
  \int_0^l \phi_\total^{}(\bs{r}, \bs{v}, l - l') S(\bs{r}, \bs{v}, l') \diff l',
\end{equation*}
we obtain
\begin{equation*}
  \Lapl \!\left[ \int_0^\infty \sigma_\total^{}(\bs{r}, \bs{v}, l) \,\xi(\bs{r}, \bs{v}, t, l) \diff l \right]
  =
  \frac{1}{v} \Lapl\phi_\total^{} \!\left( \bs{r}, \bs{v}, \frac{s}{v} + \bs\varOmega \cdot \gradr \right) \Lapl\psi(\bs{r}, \bs{v}, s),
\end{equation*}
or, equivalently,
\begin{equation*}
  \int_0^\infty \sigma_\total^{}(\bs{r}, \bs{v}, l) \,\xi(\bs{r}, \bs{v}, t, l) \diff l
  =
  \int_0^{v t} \phi_\total^{}(\bs{r}, \bs{v}, l) \,\psi \!\left( \bs{r} - \bs\varOmega l, \bs{v}, t - \dfrac{l}{v} \right) \diff l.
\end{equation*}

In the same way we obtain
\begin{equation*}
  \int_0^\infty \sigma_\scatt^{}(\bs{r}, \bs{v}, l) \,\xi(\bs{r}, \bs{v}, t, l) \diff l
  =
  \int_0^{v t} \phi_\scatt^{}(\bs{r}, \bs{v}, l) \,\psi \!\left( \bs{r} - \bs\varOmega l, \bs{v}, t - \dfrac{l}{v} \right) \diff l,
\end{equation*}
where the memory kernel $\phi_\scatt^{}$ is determined by Eq.\,\eqref{eq:MemoryKernel} with $p_\scatt^{}$ instead of $p$, while $p_\scatt^{}$ is determined by Eqs.\,\eqref{eq:PDF} with $\sigma_\scatt^{}$ instead of $\sigma$.

As a result Eq.\,\eqref{eq:PDEDensity} is recast as the \emph{linear Boltzmann-like equation}
\begin{multline}
  \pd_t^{} \psi
  + \bs{v} \cdot \gradr\psi
  =
  \int_{\spaceV}^{} K(\bs{r}, \bs{v}, \bs{v}') \left[ v' \int_0^{v' t} \phi_\scatt^{}(\bs{r}, \bs{v}', l) \,\psi \!\left( \bs{r} - \bs\varOmega' l, \bs{v}', t - \dfrac{l}{v'} \right) \diff l \right] \diff\bs{v}'\\
  - v \int_0^{v t} \phi_\total^{}(\bs{r}, \bs{v}, l) \,\psi \!\left( \bs{r} - \bs\varOmega l, \bs{v}, t - \dfrac{l}{v} \right) \diff l
  + F(\bs{r}, \bs{v}, t),
  \tag*{\eqref{eq:BoltzLikeEq}}
\end{multline}
or concisely
\begin{equation}
  \label{eq:BoltzLikeEq}
  \pd_t^{} \psi
  + \bs{v} \cdot \gradr\psi
  =
  \left( \mcal{K} \mcal{M}_\scatt^{} - \mcal{M} \right) \psi
  + F,
\end{equation}
where the scattering operator $\mcal{K}$ is given by Eq.\,\eqref{eq:ScattOperator}, and the memory operator is given by
\begin{multline}
  \label{eq:MemoryOperator}
  \mcal{M} \psi(\bs{r}, \bs{v}, t)
  =
  v \int_0^{v t} \phi_\total^{}(\bs{r}, \bs{v}, l) \,\psi \!\left( \bs{r} - \bs\varOmega l, \bs{v}, t - \dfrac{l}{v} \right) \diff l\\
  \equiv
  \int_0^t \varPhi_\total^{}(\bs{r}, \bs{v}, t') \,\psi(\bs{r} - \bs{v} t', \bs{v}, t - t') \diff t'
\end{multline}
with
\begin{equation*}
  v^2 \phi_\total^{}(\bs{r}, \bs{v}, l)
  =
  \varPhi_\total^{}(\bs{r}, \bs{v}, t),
\end{equation*}
$l = v t$,
the operator $\mcal{M}_\scatt^{}$ is determined by Eq.\,\eqref{eq:MemoryOperator} with $\phi_\scatt^{}$ and $\varPhi_\scatt^{}$ instead of $\phi_\total^{}$ and $\varPhi_\total^{}$, respectively.
The initial condition~\eqref{eq:InitCondDensADelta} results in the initial condition
\begin{equation}
  \label{eq:InitCondBoltzEq}
  \left. \psi \right|_{t=0}^{}
  =
  \chi(\bs{r}, \bs{v}).
\end{equation}
Note that an equation similar to Eq.\,\eqref{eq:BoltzLikeEq} (without absorption and the source term) was obtained in Ref.~\cite{FriedrichEtAl:2006_PRE}, see also~\cite{FriedrichEtAl:2006_PRL}.

If the extinction and scattering coefficients depend only on velocity, \ie, $\sigma_\total^{}(\bs{r}, \bs{v}, l) \equiv \sigma_\total^{}(\bs{v})$ and $\sigma_\scatt^{}(\bs{r}, \bs{v}, l) \equiv \sigma_\scatt^{}(\bs{v})$, then the memory kernels are $\phi(\bs{r}, \bs{v}, l) = \sigma_\total^{}(\bs{v}) \,\delta(l)$ and $\phi_\scatt^{}(\bs{r}, \bs{v}, l) = \sigma_\scatt^{}(\bs{v}) \,\delta(l)$. In this case $\mcal{M} \psi = \sigma_\total^{}(\bs{v}) v \psi$ and $\mcal{M}_\scatt^{} \psi = \sigma_\scatt^{}(\bs{v}) v \psi$. Hence the Boltzmann-like equation~\eqref{eq:BoltzLikeEq} becomes the \emph{ordinary linear Boltzmann equation} (without the acceleration term)
\begin{equation}
  \label{eq:BoltzEq}
  \pd_t^{} \psi
  + \bs{v} \cdot \gradr\psi
  =
  \int_{\spaceV}^{} K(\bs{r}, \bs{v}, \bs{v}') \sigma_\scatt^{}(\bs{v}') v' \psi(\bs{r}, \bs{v}', t) \diff\bs{v}'
  - \sigma_\total^{}(\bs{v}) v \psi
  + F(\bs{r}, \bs{v}, t).
\end{equation}

One can derive integral equations for the densities $\eta$ and $\psi$, see Appendix~\ref{sec:IntegralEquations}.

\section{Asymptotic solution for small mean free paths}
\label{sec:AsySolution}

The Boltzmann-like equation~\eqref{eq:BoltzLikeEq} describes transport with an arbitrary distribution of the free path. But here we consider distributions with at least first and second finite moments.

\subsection{Assumptions}

We suppose that the ``space'' of velocities $\spaceV$ is bounded and rotationally invariant, and $\bs{0} \notin \spaceV$, \ie, all particles have nonzero velocities.
We suppose also that the medium is isotropic and the kernel has the form
\begin{equation}
  \label{eq:KernelIsotropic}
  K(\bs{r}, \bs{v}, \bs{v}')
  \equiv
  K(\bs{r}, \bs\varOmega \cdot \bs\varOmega', v, v')
\end{equation}
where $\bs\varOmega = \bs{v} / v$, $v = |\bs{v}|$, and the same for $\bs\varOmega'$.
Besides, the kernel is supposed to be bounded from below and above:
\begin{equation*}
  0 < K_{\text{min}} \leq K \leq K_{\text{max}} < \infty.
\end{equation*}
This implies that $\int_\spaceV \int_\spaceV |K(\cdot, \bs{v}, \bs{v}')|^2 \diff\bs{v}' \diff\bs{v} < \infty$.

We suppose here that the scattering coefficient is equal to $\sigma_\scatt^{}(\bs{r}, \bs{v}, l) = \omega(\bs{r}) \sigma_\total^{}(\bs{r}, \bs{v}, l)$, where $\omega$ is the scattering albedo. Then the Boltzmann-like equation~\eqref{eq:BoltzLikeEq} takes the form
\begin{equation}
  \label{eq:BoltzLikeEqOmega}
  \pd_t^{} \psi
  + \bs{v} \cdot \gradr\psi
  =
  \left( \omega \mcal{K} - \Id \right) \mcal{M} \psi
  + F.
\end{equation}

If the mean free path is small of order $\varepsilon$, where $\varepsilon$ a small parameter, the extinction coefficient is represented as
\begin{equation}
  \label{eq:ExtincCoeffEps}
  \sigma_\total^{}(v, l)
  =
  \frac{1}{\varepsilon} \,\bar{\sigma}_\total^{} \!\left( v, \frac{l}{\varepsilon} \right),
\end{equation}
see Appendix~\ref{sec:MemoryKernelAsymptoticExpansion}.
This implies representation for the memory kernel
\begin{equation}
  \label{eq:MemoryKernelEps}
  \phi(v, l)
  =
  \frac{1}{\varepsilon^2} \,\bar{\phi} \!\left( v, \frac{l}{\varepsilon} \right).
\end{equation}
We suppose that the scattering albedo is given by
\begin{equation}
  \label{eq:OmegaEps}
  \omega
  =
  1 - \varepsilon^2 \alpha(\bs{r}).
\end{equation}
This means that scattering dominates the transport, while absorption is small of order $\varepsilon^2$ compared to scattering.
Finally, we suppose that the source term is small of order $\varepsilon$ and represented by
\begin{equation}
  \label{eq:SourceEps}
  F = \varepsilon \bar{F}.
\end{equation}

\subsection{The equation and representation of the solution}

Taking into account all the assumptions we recast the Boltzmann-like equation as
\begin{multline}
  \label{eq:BoltzLikeEqEps}
  \varepsilon \pd_t^{} \psi
  + \varepsilon \bs{v} \cdot \gradr\psi
  =
  \left[ (1 - \varepsilon^2 \alpha) \mcal{K} - \Id \right] \left[ v \int_0^{v t} \frac{1}{\varepsilon} \bar{\phi} \!\left( v, \frac{l}{\varepsilon} \right) \psi \!\left( \bs{r} - \bs\varOmega l, \bs{v}, t - \dfrac{l}{v} \right) \diff l \right]\\
  + \varepsilon^2 \bar{F}(\bs{r}, \bs{v}, t)
\end{multline}
To find the asymptotic solution we represent the density in the form
\begin{equation*}
  \psi(\bs{r}, \bs{v}, t)
  =
  \psi^\outersol(\bs{r}, \bs{v}, t^\outersol)
  + \psi^\innersol(\bs{r}, \bs{v}, t^\innersol)
\end{equation*}
with $t^\outersol = \varepsilon t$ and $t^\innersol = t / \varepsilon$, where $\psi^\outersol$ and $\psi^\innersol$ are outer and inner solutions, respectively. The outer solution approximates the solution in the interior of the domain, while the inner solution approximates it in the initial layer.

\subsection{The outer expansion}

Substituting $\psi^\outersol$ into Eq.~\eqref{eq:BoltzLikeEqEps} yields
\begin{multline}
  \label{eq:BoltzLikeEqOuter}
  \varepsilon^2 \pd_{t^\outersol}^{} \psi^\outersol
  + \varepsilon \bs{v} \cdot \gradr\psi^\outersol\\
  =
  \left[ (1 - \varepsilon^2 \alpha) \mcal{K} - \Id \right] \left[ v \int_0^{v t^\outersol/\varepsilon} \frac{1}{\varepsilon} \bar{\phi} \!\left( v, \frac{l}{\varepsilon} \right) \psi^\outersol \!\!\left( \bs{r} - \bs\varOmega l, \bs{v}, t^\outersol - \dfrac{\varepsilon l}{v} \right) \diff l \right]
  + \varepsilon^2 \bar{F} \!\left( \bs{r}, \bs{v}, \frac{t^\outersol}{\varepsilon} \right).
\end{multline}
[This equation differs from Eq.~\eqref{eq:BoltzLikeEqEps} only in the first term on the left-hand side.]
We assume that $\psi^\outersol$ has the asymptotic expansion
\begin{equation*}
  \psi^\outersol
  \sim
  \psi_0^\outersol
  + \varepsilon \psi_1^\outersol
  + \varepsilon^2 \psi_2^\outersol
  + \ldots{}
\end{equation*}
Besides, we have the asymptotic expansion (see Appendix~\ref{sec:MemoryKernelAsymptoticExpansion})
\begin{equation}
  \label{eq:MemoryKernelAsyExpansion}
  \frac{1}{\varepsilon} \,\bar{\phi} \!\left( v, \frac{l}{\varepsilon} \right)
  \sim
  \phi_0^{} \delta(l)
  + \phi_1^{} \delta'(l) \varepsilon
  + \phi_2^{} \delta''(l) \varepsilon^2
  + \ldots{}
\end{equation}
with
\begin{equation}
  \label{eq:CoeffMemoryKernelAsy}
  \phi_0^{}
  \equiv
  \phi_0^{}(v)
  =
  \frac{1}{\average{\overline{l}}}
  \quad\text{and}\quad
  \phi_1^{}
  \equiv
  \phi_1^{}(v)
  =
  \frac{\average{\overline{l}^2}}{2 \average{\overline{l}}^2} - 1.
\end{equation}
Substituting the asymptotic expansions into Eq.~\eqref{eq:BoltzLikeEqOuter} and equating coefficients of like powers of $\varepsilon$ yields the following equations:
\begin{subequations}
  \label{eq:EqEps}
  \begin{align}
    \label{eq:EqEpsZero}
    & \varepsilon^0:
    \qquad
    \left( \mcal{K} - \Id \right) \left( \phi_0^{} v \psi_0^\outersol \right)
    = 0,\\
    \label{eq:EqEpsOne}
    & \varepsilon^1:
    \qquad
    \left( \mcal{K} - \Id \right) \left( \phi_0^{} v \psi_1^\outersol \right)
    =
    \bs\varOmega \cdot \left[
      v \gradr \psi_0^\outersol
      + \left( \Id - \mcal{K}_1^{} \right) \left( \phi_1^{} v \gradr \psi_0^\outersol \right)
    \right],\\
    & \varepsilon^2:
    \qquad
    \left( \mcal{K} - \Id \right) \left( \phi_0^{} v \psi_2^\outersol \right)
    =
    \pd_{t^\outersol}^{} \psi_0^\outersol
    + v \bs\varOmega \cdot \gradr \psi_1^\outersol\notag\\
    & \hspace*{12.2em} 
    + \left( \Id - \mcal{K} \right) \left[
      \phi_1^{} \pd_{t^\outersol} \psi_0^\outersol
      + \phi_1^{} \bs{v} \cdot \gradr \psi_1^\outersol
      + \phi_2^{} \frac{1}{v} (\bs{v} \cdot \gradr)^2 \psi_0^\outersol
    \right]\notag\\
    \label{eq:EqEpsTwo}
    & \hspace*{12.2em}
    + \mcal{K} \left( \alpha \phi_0^{} v \psi_0^\outersol \right)
    - \bar{F} \bigg( \bs{r}, \bs{v}, \frac{t^\outersol}{\varepsilon} \bigg).
  \end{align}
\end{subequations}

Eq.~\eqref{eq:EqEpsZero} implies (see Appendix~\ref{sec:FredholmEq})
\begin{equation}
  \label{eq:DensThroughDens}
  \psi_0^\outersol(\bs{r}, v, t^\outersol)
  =
  \frac{\varPsi(\bs{r}, v)}{\phi_0^{}(v) v} \rho(\bs{r}, t^\outersol),
\end{equation}
where $\varPsi$ is a positive solution of the equation $\left( \Id - \mcal{K} \right) \varPsi = 0$, and $\rho$ is not yet determined.

\textbf{Remark:} Instead of using the isotropic kernel~\eqref{eq:KernelIsotropic} one may use a general kernel $K(\bs{r}, \bs{v}, \bs{v}')$, impose a condition $\varPsi \equiv 1$ and take $\phi_0^{}(v) v = \lambda_0^{} \equiv \const$ as in Ref.~\cite{HillenOthmer:2000}. In this case $\psi_0^\outersol(\bs{r}, v, t) = \rho(\bs{r}, t) / \lambda_0^{}$.

Eq.~\eqref{eq:EqEpsOne} yields the solution (see Appendix~\ref{sec:FredholmEq})
\begin{equation*}
  \psi_1^\outersol(\bs{r}, \bs{v}, t^\outersol)
  =
  \frac{1}{\phi_0^{} v}
  \left[
    c \varPsi
    - \bs\varOmega \cdot \gradr \left( A \rho \right)
  \right]
  \equiv
  \frac{1}{\phi_0^{} v}
  \left[
    c \varPsi
    - A \bs\varOmega \cdot \gradr \rho
    - \left( \bs\varOmega \cdot \gradr A \right) \rho
  \right],
\end{equation*}
where $c \equiv c(\bs{r}, t^\outersol)$ is an arbitrary function, and
\begin{equation*}
  A(\bs{r}, v)
  =
  \left( \Id - \mcal{K}_1^{} \right)^{-1} \!\left( \frac{1}{\phi_0^{}} \varPsi \right) + \frac{\phi_1^{}}{\phi_0^{}} \varPsi.
\end{equation*}
Substituting $\psi_1^\outersol$ into Eq.~\eqref{eq:EqEpsTwo} and using the solvability condition (see Appendix~\ref{sec:FredholmEq}) yields the equation
\begin{equation}
  \label{eq:EqForDensityRho}
  C \pd_{t^\outersol}^{} \rho
  - \diverg \left( B \gradr \rho \right)
  - \gradr B \cdot \gradr \rho
  + \left( - \Delta B +  \int \alpha \varPsi v^2 \diff v \right) \rho
  =
  \frac{1}{4 \pi} \int_\spaceV \bar{F} \!\left( \bs{r}, \bs{v}, \frac{t^\outersol}{\varepsilon} \right) \diff \bs{v},
\end{equation}
with
\begin{equation*}
  C(\bs{r})
  =
  \int \frac{\varPsi v}{\phi_0^{}} \diff v
\end{equation*}
and
\begin{equation*}
  B(\bs{r})
  =
  \frac{1}{3} \int \frac{v^2}{\phi_0^{}(v)} A(\bs{r}, v) \diff v.
\end{equation*}
Hence $\psi_0^\outersol$ is given by Eq.~\eqref{eq:DensThroughDens}, where $\rho$ satisfies Eq.~\eqref{eq:EqForDensityRho}.

\subsection{The inner expansion}

Since the outer solution $\psi^\outersol$ satisfies the Boltzmann-like equation, the inner solution $\psi^\innersol$ should satisfy the homogeneous Boltzmann-like equation.
Substituting $\psi^\innersol$ into  Eq.~\eqref{eq:BoltzLikeEqEps} with $\bar{F} \equiv 0$ yields
\begin{equation}
  \label{eq:BoltzLikeEqInner}
  \pd_{t^\innersol} \psi^\innersol
  + \varepsilon \bs{v} \cdot \gradr\psi^\innersol
  =
  \left[ (1 - \varepsilon^2 \alpha) \mcal{K} - \Id \right] \left[ v \int_0^{\varepsilon v t^\innersol} \frac{1}{\varepsilon} \bar{\phi} \!\left( v, \frac{l}{\varepsilon} \right) \psi^\innersol \!\left( \bs{r} - \bs\varOmega l, \bs{v}, t^\innersol - \dfrac{l}{\varepsilon v} \right) \diff l \right].
\end{equation}
We assume that $\psi^\innersol$ has the asymptotic expansion
\begin{equation*}
  \psi^\innersol
  \sim
  \psi_0^\innersol
  + \varepsilon \psi_1^\innersol
  + \ldots{}
\end{equation*}
Substituting this expansion and \eqref{eq:MemoryKernelAsyExpansion} into Eq.~\eqref{eq:BoltzLikeEqInner} and equating coefficients of $\varepsilon^0$ yields the equation
\begin{equation*}
  \pd_{t^\innersol} \psi_0^\innersol
  =
  \left( \mcal{K} - \Id \right) \left( \phi_0^{} v \psi_0^\innersol \right),
\end{equation*}
which is in fact an ordinary differential equation. Its solution is
\begin{equation}
  \label{eq:SolInner}
  \psi_0^\innersol(\bs{r}, \bs{v}, t^\innersol)
  =
  \frac{1}{\phi_0^{} v} \e^{\phi_0^{} v t^\innersol \left( \mcal{K} - \Id \right)} \left( \phi_0^{} v \left. \psi_0^\innersol \right|_{t^\innersol = 0} \right).
\end{equation}

\subsection{Initial conditions}

The initial distribution $\chi$ does not depend on $\varepsilon$, therefore we have the initial condition
\begin{equation*}
  \left. \left[ \psi_0^\outersol
  + \psi_0^\innersol \right] \!\right|_{t = 0}^{}
  =
  \chi(\bs{r}, \bs{v}).
\end{equation*}
Due to properties of the scattering operator $\mcal{K}$ (see Appendix~\ref{sec:FredholmEq}), in order to ensure that $\psi_0^\innersol \to 0$ as $t^\innersol \to \infty$, the initial ``value'' $\phi_0^{} v \left. \psi_0^\innersol \right|_{t^\innersol = 0}$ in Eq.~\eqref{eq:SolInner} must be orthogonal to $\Psi$ in $L_2^{}(\spaceV)$.
This results in the initial conditions
\begin{equation*}
  \left. \psi_0^\outersol \right|_{t=0}^{}
  =
  \frac{\varPsi(\bs{r}, v)}{\phi_0^{}(v) v} \rho \big|_{t=0}^{}
\end{equation*}
with
\begin{equation}
  \label{eq:InitCondDensRho}
  \rho \big|_{t=0}^{}
  =
  \frac{\left< \phi_0^{} v \chi, \varPsi \right>_{L_2^{}(\spaceV)}}{\left\| \varPsi \right\|_{L_2^{}(\spaceV)}^2}
\end{equation}
and
\begin{equation}
  \label{eq:InitCondDensInner}
  \psi_0^\innersol \Big|_{t=0}^{}
  =
  \chi(\bs{r}, \bs{v})
  - \psi_0^\outersol \Big|_{t=0}^{}.
\end{equation}

\subsection{Summary}

The asymptotic solution of the Boltzmann-like equation~\eqref{eq:BoltzLikeEqOmega} with the extinction coefficient~\eqref{eq:ExtincCoeffEps} [this implies the memory kernel~\eqref{eq:MemoryKernelEps}], scattering albedo~\eqref{eq:OmegaEps} and source term~\eqref{eq:SourceEps} is given by 
\begin{equation*}
  \psi_0^{}(\bs{r}, \bs{v}, t)
  =
  \psi_0^\outersol(\bs{r}, \bs{v}, t^\outersol)
  + \psi_0^\innersol(\bs{r}, \bs{v}, t^\innersol),
\end{equation*}
with $t^\outersol = \varepsilon t$ and $t^\innersol = t / \varepsilon$, where $\psi_0^\outersol$ is given by Eq.~\eqref{eq:DensThroughDens} with $\rho$ obeying Eq.~\eqref{eq:EqForDensityRho}, and
$\psi_0^\innersol$ is given by Eq.~\eqref{eq:SolInner}.
The initial condition~\eqref{eq:InitCondBoltzEq} results in the initial conditions~\eqref{eq:InitCondDensRho} and \eqref{eq:InitCondDensInner} for $\rho$ and $\psi_0^\innersol$, respectively.
Note that $\psi_0^\innersol$ decays exponentially as time increases.

\section{The diffusion approximation to the one-speed Boltzmann-like equation}
\label{sec:DiffApprox}

\subsection{The asymptotic solution}

The asymptotic solution, obtained in the previous section, becomes a diffusion approximation in the case of the one-speed equation.
Since the velocity $v$ is constant, the density can be given by
\begin{equation*}
  \psi
  \equiv
  \psi(\bs{r}, \bs\varOmega, t).
\end{equation*}
Then the scattering kernel is given by
\begin{equation*}
  K(\bs{r}, \bs\varOmega, \bs\varOmega')
  \equiv
  K(\bs{r}, \bs\varOmega \cdot \bs\varOmega'),
\end{equation*}
and the scattering operator becomes
\begin{equation*}
  \mcal{K} \psi(\bs{r}, \bs\varOmega, t)
  =
  \int_\sphere K(\bs{r}, \bs\varOmega \cdot \bs\varOmega') \psi(\bs{r}, \bs\varOmega', t) \diff\bs\varOmega'.
\end{equation*}
The Boltzmann-like equation takes the form of the one-speed equation
\begin{equation}
  \label{eq:BoltzLikeEqOneSpeed}
  \pd_t^{} \psi
  + v \bs\varOmega \cdot \gradr\psi
  =
  \left( \omega \mcal{K} - \Id \right) \left[ v \int_0^{v t} \phi_\total^{}(\bs{r}, \bs\varOmega, l) \,\psi \!\left( \bs{r} - \bs\varOmega l, \bs\varOmega, t - \dfrac{l}{v} \right) \diff l \right] \diff \bs\varOmega
  + F(\bs{r}, \bs\varOmega, t),
\end{equation}
where $\omega$ is the scattering albedo.
The initial condition is
\begin{equation}
  \label{eq:InitCondBoltzEqOneSpeed}
  \left. \psi \right|_{t=0}^{}
  =
  \chi(\bs{r}, \bs\varOmega).
\end{equation}

Eq.~\eqref{eq:BoltzLikeEqEps} becomes
\begin{multline*}
  \varepsilon \pd_t^{} \psi
  + \varepsilon v \bs\varOmega \cdot \gradr\psi\\
  =
  \left[ (1 - \varepsilon^2 \alpha) \mcal{K} - \Id \right] \!\left[ v \int_0^{v t} \frac{1}{\varepsilon} \bar{\phi} \!\left( \frac{l}{\varepsilon} \right) \psi \!\left( \bs{r} - \bs\varOmega l, \bs\varOmega, t - \dfrac{l}{v} \right) \diff l \right]
  + \varepsilon^2 \bar{F}(\bs{r}, \bs\varOmega, t).
\end{multline*}
We represent the density as
\begin{equation*}
  \psi(\bs{r}, \bs\varOmega, t)
  =
  \psi^\outersol(\bs{r}, \bs\varOmega, t^\outersol)
  + \psi^\innersol(\bs{r}, \bs\varOmega, t^\innersol)
\end{equation*}
with $t^\outersol = \varepsilon t$ and $t^\innersol = t / \varepsilon$, where $\psi^\outersol$ and $\psi^\innersol$ are the outer and inner solutions, respectively.

Eqs.~\eqref{eq:EqEps} become
\begin{subequations}
  \label{eq:EqEpsOneSpeed}
  \begin{align}
    \label{eq:EqEpsOneSpeedZero}
    & \varepsilon^0:
    \qquad
    \phantom{\phi_0^{} v} \left( \mcal{K} - \Id \right) \psi_0^\outersol
    = 0,\\
    \label{eq:EqEpsOneSpeedOne}
    & \varepsilon^1:
    \qquad
    \phi_0^{} v \left( \mcal{K} - \Id \right) \psi_1^\outersol
    =
    \bs\varOmega \cdot \left[ 1 + \left( 1 - K_1^{} \right) \phi_1^{} \right] v \gradr \psi_0^\outersol,\\
    & \varepsilon^2:
    \qquad
    \phi_0^{} v \left( \mcal{K} - \Id \right) \psi_2^\outersol
    =
    \pd_{t^\outersol}^{} \psi_0^\outersol
    + v \bs\varOmega \cdot \gradr \psi_1^\outersol
    + \left( \Id - \mcal{K} \right) \left( \ldots \right)\notag\\
    \label{eq:EqEpsOneSpeedTwo}
    & \hspace*{11.3em} 
    + \mcal{K} \!\left( \alpha \phi_0^{} v \psi_0^\outersol \right)
    - \bar{F} \bigg( \bs{r}, \bs\varOmega, \frac{t^\outersol}{\varepsilon} \bigg),
  \end{align}
\end{subequations}
with $\phi_i^{}$ given by Eq.~\eqref{eq:CoeffMemoryKernelAsy}.
The coefficient $K_1^{} \equiv K_1^{}(\bs{r})$ is the same as in Appendix~\ref{sec:FredholmEq}, but here it does not depend on $v$ and $v'$.
Eq.~\eqref{eq:EqEpsOneSpeedZero} implies (see Appendix~\ref{sec:FredholmEq}, note that $\varPsi = 1$)
\begin{equation*}
  \psi_0^\outersol(\bs{r}, \bs\varOmega, t^\outersol)
  \equiv
  \psi_0^\outersol(\bs{r}, t^\outersol),
\end{equation*}
\ie, $\psi_0^\outersol$ does not depend on $\bs\varOmega$.
Eq.~\eqref{eq:EqEpsOneSpeedOne} yields the solution (see Appendix~\ref{sec:FredholmEq})
\begin{equation*}
  \psi_1^\outersol(\bs{r}, \bs\varOmega, t^\outersol)
  =
  c - A \bs\varOmega \cdot \gradr \psi_0^\outersol,
\end{equation*}
where $c \equiv c(\bs{r}, t^\outersol)$ is an arbitrary function, and
\begin{equation*}
  A(\bs{r})
  =
  \frac{1}{\phi_0^{}} \left[ \frac{1}{1 - K_1^{}(\bs{r})} + \phi_1^{} \right]
  \equiv
  \average{\overline{l}} \!\left[ \frac{\average{\overline{l}^2}}{2 \average{\overline{l}}^2} + \frac{K_1^{}(\bs{r})}{1 - K_1^{}(\bs{r})} \right].
\end{equation*}
Substituting $\psi_1^\outersol$ into Eq.~\eqref{eq:EqEpsOneSpeedTwo} and using the solvability condition yields the diffusion equation for $\psi_0^\outersol$
\begin{equation}
  \label{eq:DiffEqDensOuter}
  \pd_{t^\outersol}^{} \psi_0^\outersol
  - \diverg \left( D^\outersol \gradr \psi_0^\outersol \right)
  + \kappa^\outersol \psi_0^\outersol
  =
  \frac{1}{4 \pi} f^\outersol(\bs{r}, t^\outersol)
\end{equation}
with
\begin{equation*}
  D^\outersol(\bs{r})
  =
  \frac{v \average{\overline{l}}}{3} \!\left[ \frac{\average{\overline{l}^2}}{2 \average{\overline{l}}^2} + \frac{K_1^{}(\bs{r})}{1 - K_1^{}(\bs{r})} \right],
  \quad
  \kappa^\outersol
  =
  \frac{\alpha v}{\average{\overline{l}}}
  \quad\text{and}\quad
  f^\outersol(\bs{r}, t^\outersol)
  =
  \int_\sphere \bar{F} \!\left( \bs{r}, \bs\varOmega, \frac{t^\outersol}{\varepsilon} \right) \diff\bs\varOmega.
\end{equation*}
The initial condition~\eqref{eq:InitCondBoltzEqOneSpeed} yields the initial condition
\begin{equation}
  \label{eq:InitCondDensOuterOneSpeed}
  \left. \psi_0^\outersol \right|_{t=0}^{}
  =
  \frac{1}{4 \pi} \int_\sphere \chi(\bs{r}, \bs\varOmega) \diff \bs\varOmega.
\end{equation}

The term $\psi_0^\innersol$ obeys the equation
\begin{equation*}
  \pd_{t^\innersol} \psi_0^\innersol
  =
  \phi_0^{} v \left( \mcal{K} - \Id \right) \psi_0^\innersol,
\end{equation*}
which implies the solution
\begin{equation}
  \label{eq:SolInnerOneSpeed}
  \psi_0^\innersol(\bs{r}, \bs{v}, t^\innersol)
  =
  \e^{\phi_0^{} v t^\innersol \left( \mcal{K} - \Id \right)} \!\left. \psi_0^\innersol \right|_{t^\innersol = 0}.
\end{equation}
The initial condition is
\begin{equation}
  \label{eq:InitCondDensInnerOneSpeed}
  \psi_0^\innersol \Big|_{t=0}^{}
  =
  \chi(\bs{r}, \bs\varOmega)
  - \frac{1}{4 \pi} \int_\sphere \chi(\bs{r}, \bs\varOmega') \diff \bs\varOmega'.
\end{equation}

In summary,
the asymptotic solution of the one-speed Boltzmann-like equation~\eqref{eq:BoltzLikeEqOneSpeed}, which satisfies the initial condition~\eqref{eq:InitCondBoltzEqOneSpeed}, is given by 
\begin{equation*}
  \psi_0^{}(\bs{r}, \bs\varOmega, t)
  =
  \psi_0^\outersol(\bs{r}, \bs\varOmega, t^\outersol)
  + \psi_0^\innersol(\bs{r}, \bs\varOmega, t^\innersol),
\end{equation*}
where $\psi_0^\outersol$ satisfies Eq.~\eqref{eq:DiffEqDensOuter} and the initial condition~\eqref{eq:InitCondDensOuterOneSpeed}, and
$\psi_0^\innersol$ is given by Eq.~\eqref{eq:SolInnerOneSpeed} with the initial condition~\eqref{eq:InitCondDensInnerOneSpeed}.
The term $\psi_0^\innersol$ decays exponentially as time increases.

\subsection{The diffusion approximation}

Multiplying Eq.~\eqref{eq:DiffEqDensOuter} by $4 \pi$ and $\varepsilon$ yields the diffusion equation
\begin{equation}
  \label{eq:DiffEq}
  \pd_t^{} u
  - \diverg \left( D \gradr u \right)
  + \kappa u
  =
  f(\bs{r}, t)
\end{equation}
for the density of the particles
\begin{equation*}
  u(\bs{r}, t)
  =
  \int_\sphere^{} \psi_0^\outersol(\bs{r}, t^\outersol) \diff\bs\varOmega
  \equiv
  4 \pi \psi_0^\outersol(\bs{r}, t^\outersol),
\end{equation*}
where
\begin{equation*}
  D(\bs{r})
  =
  \frac{\average{l} v}{3} \!\left[ \frac{\average{l^2}}{2 \average{l}^2} + \frac{K_1^{}(\bs{r})}{1 - K_1^{}(\bs{r})} \right],
  \quad
  \kappa
  =
  \frac{(1 - \omega) v}{\average{l}}
  \quad\text{and}\quad
  f(\bs{r}, t)
  =
  \int_\sphere F(\bs{r}, \bs\varOmega, t) \diff\bs\varOmega.
\end{equation*}
The initial condition~\eqref{eq:InitCondDensOuterOneSpeed} yields the initial condition
\begin{equation}
  \label{eq:InitCondDiffEq}
  \left. u \right|_{t=0}^{}
  =
  \int_\sphere \chi(\bs{r}, \bs\varOmega) \diff \bs\varOmega.
\end{equation}

If $\sigma_\total^{} = \const$, the distribution of the path length is exponential, \ie, $p(l) = \sigma_\total^{} \e^{-\sigma_\total^{} l}$, and $\phi(l) = \sigma \delta(l)$. In this case the one-speed Boltzmann-like equation~\eqref{eq:BoltzLikeEqOneSpeed} becomes the ordinary one-speed linear Boltzmann equation
\begin{equation}
  \label{eq:BoltzEqOneSpeed}
  \pd_t^{} \psi
  + v \bs\varOmega \cdot \gradr\psi
  =
  \left( \omega \mcal{K} - \Id \right) \left( \sigma_\total^{} v \psi \right)
  + F(\bs{r}, \bs\varOmega, t),
\end{equation}
the moments are $\average{l} = 1 / \sigma_\total^{}$ and $\average{l^2} = 2 / \sigma_\total^2$, and we obtain
\begin{equation}
  \label{eq:DiffAbsorpCoeffDiffApprox}
  D(\bs{r})
  =
  \frac{v}{3 \sigma_\total^{} \left[ 1 - K_1^{}(\bs{r}) \right]}
  \quad\text{and}\quad
  \kappa
  =
  (1 - \omega) \sigma_\total^{} v.
\end{equation}
The diffusion approximation to the Boltzmann equation~\eqref{eq:BoltzEqOneSpeed} can be obtained by the method of spherical harmonics~\cite{DuderstadtMartin:1979, Modest:2013}. The latter yields the diffusion coefficient $D = v / \left[ 3 \sigma_\total^{} (1 - \omega K_1^{}) \right]$, which differs from the diffusion coefficient~\eqref{eq:DiffAbsorpCoeffDiffApprox} by $O(\varepsilon^2)$, since $\omega = 1 - O(\varepsilon^2)$.
The absorption coefficient and initial condition, obtained by the method of spherical harmonics, coincide with the absorption coefficient~\eqref{eq:DiffAbsorpCoeffDiffApprox} and the initial condition~\eqref{eq:InitCondDiffEq}, respectively.

Note that, if transport is considered in the two-dimensional space $\mbb{R}^2$, the diffusion coefficient is
\begin{equation*}
  D(\bs{r})
  =
  \frac{\average{l} v}{2} \!\left[ \frac{\average{l^2}}{2 \average{l}^2} + \frac{K_1^{}(\bs{r})}{1 - K_1^{}(\bs{r})} \right].
\end{equation*}
This coincides with the diffusion coefficient obtained in Ref.\,\cite{ShaebaniEtAl:2014}: see Eq.~(3) in Ref.\,\cite{ShaebaniEtAl:2014} with $p = 0$ and $\mcal{R} = K_1^{}$.

\section*{Acknowledgments}

This work was partially supported by the RFBR grant No.\,14-01-00334 and by State commission contract in scientific area (project part).

\appendix

\section{The phase space density, free path distribution and related functions}
\label{sec:DensityOnCharacteristic}

The behaviour of the phase space density $\xi$ along the characteristics of the first-order partial differential equation~\eqref{eq:PDEDensA} can be completely described through the dependence on the variable $l$. Therefore, if the dependence of $\xi$ on position $\bs{r}$, velocity $\bs{v}$ and time $t$ is omitted this behaviour is described by the ordinary differential equation
\begin{equation*}
  \xi' + \sigma_\total^{} \xi = 0,
  \quad
  l > 0,
\end{equation*}
where $\xi \equiv \xi(l)$ and $\sigma_\total^{} \equiv \sigma_\total^{}(l)$ is the extinction coefficient.
The solution of this equation is given by
\begin{equation*}
  \xi(l)
  =
  S(l) \xi(0),
\end{equation*}
where
\begin{equation*}
  S(l)
  =
  \exp \left\{ - \int_0^l \sigma_\total^{}(l') \diff l' \right\}
  =
  1 - \int_0^l p(l') \diff l'
  \equiv
  \int_l^\infty p(l') \diff l'
\end{equation*}
is the survival probability, \ie, the probability that a free path of the particle is not less than $l$, and
\begin{equation*}
  p(l)
  =
  \sigma_\total^{}(l) \exp \left\{ - \int_0^l \sigma_\total^{}(l') \diff l' \right\}
  \equiv
  \sigma_\total^{}(l) S(l)
\end{equation*}
is the probability density function of a free path.

We define the memory kernel $\phi$ by~\cite{KenkreEtAl:1973}
\begin{equation*}
  \Lapl\phi(\lambda)
  \defin
  \frac{\Lapl{p}(\lambda)}{\Lapl{S}(\lambda)}
  \equiv
  \frac{\lambda \Lapl{p}(\lambda)}{1 - \Lapl{p}(\lambda)},
\end{equation*}
where
\begin{equation*}
  \Lapl{f}(\lambda)
  =
  \int_0^\infty \e^{-\lambda l} f(l) \diff l
\end{equation*}
is the Laplace transform.
This means that
\begin{equation*}
  p(l)
  =
  \int_0^l \phi(l - l') S(l') \diff l'.
\end{equation*}
If $\sigma_\total^{} \equiv \const$, \ie, the distribution is exponential, then $\phi(l) = \sigma_\total^{} \delta(l)$. This justifies the term ``memory kernel''.

The mean free path is given by
\begin{equation*}
  \average{l}
  \equiv
  \int_0^\infty l p(l) \diff l
  =
  \int_0^\infty S(l) \diff l.
\end{equation*}
The second moment is given by
\begin{equation*}
  \average{l^2}
  \equiv
  \int_0^\infty l^2 p(l) \diff l
  =
  2 \int_0^\infty l S(l) \diff l.
\end{equation*}

\section{Integral equations for the densities}
\label{sec:IntegralEquations}

Eqs.~\eqref{eq:BoundCondDensA} and \eqref{eq:IntSigmaDensA} (with the PDF $p_\scatt^{}$ instead of $p_\total^{}$, corresponding to the scattering coefficient $\sigma_\scatt^{}$) yield the integral equation for the density $\eta$
\begin{equation*}
  \eta(\bs{r}, \bs{v}, t)
  =
  \mcal{K} \left[
    \int_0^{v t} p_\scatt^{}(\bs{r}, \bs{v}, l) \,\eta \!\left( \bs{r} - \bs\varOmega l, \bs{v}, t - \dfrac{l}{v} \right) \diff l
    +
    p_\scatt^{}(\bs{r}, \bs{v}, v t) \chi(\bs{r} - \bs{v} t, \bs{v})
  \right]
  + F(\bs{r}, \bs{v}, t),
\end{equation*}
cf.~Eq.\,(2.8) in Ref.~\cite{Uchaikin:1998}, where the initial distribution $\chi$ is absent.

Integration of the Boltzmann-like equation~\eqref{eq:BoltzLikeEq} along the characteristics and the initial condition~\eqref{eq:InitCondBoltzEq} yield the integral equation
\begin{multline*}
  \psi(\bs{r}, \bs{v}, t)\\
  =
  \int_0^t \left( \mcal{K} \mcal{M}_\scatt^{} - \mcal{M}\right) \psi(\bs{r} - \bs{v} t', \bs{v}, t - t') \diff t'
  + \int_0^t F(\bs{r} - \bs{v} t', \bs{v}, t - t') \diff t'
  + \chi(\bs{r} - \bs{v} t, \bs{v}).
\end{multline*}

\section{The asymptotic expansion of the memory kernel}
\label{sec:MemoryKernelAsymptoticExpansion}

In this section the dependence of values and functions on $\bs{v}$ is omitted.

We suppose that the mean free path $\average{l}$ is small of order $\varepsilon$, where $\varepsilon$ a small parameter. To be specific, we represent it by $\average{l} = \varepsilon \average{\overline{l}}$, where $\average{\overline{l}} = O(1)$ as $\varepsilon \to 0$ [the symbol $O$ is used here in the strict sense and means ``strictly of the order of'', \ie, $f = O(g)$ means $f = O(g)$ and $f \neq o(g)$].
Then the extinction coefficient is represented as
\begin{equation*}
  \sigma_\total^{}(l)
  =
  \frac{1}{\varepsilon} \,\bar{\sigma}_\total^{} \!\left( \frac{l}{\varepsilon} \right).
\end{equation*}
This implies
\begin{equation*}
  p(l)
  =
  \frac{1}{\varepsilon} \,\bar{p} \!\left( \frac{l}{\varepsilon} \right),
  \quad
  S(l)
  =
  \bar{S} \!\left( \frac{l}{\varepsilon} \right)
  \quad\text{and}\quad
  \phi(l)
  =
  \frac{1}{\varepsilon^2} \,\bar{\phi} \!\left( \frac{l}{\varepsilon} \right)
\end{equation*}
with
\begin{equation*}
  \Lapl\bar{\phi}(\lambda)
  =
  \frac{\Lapl{\bar{p}}(\lambda)}{\Lapl{\bar{S}}(\lambda)}
  \equiv
  \frac{\lambda \Lapl{\bar{p}}(\lambda)}{1 - \Lapl{\bar{p}}(\lambda)}.
\end{equation*}
The first and second moments are given by
\begin{equation*}
  \average{\overline{l}}
  \equiv
  \int_0^\infty l \bar{p}(l) \diff l
\end{equation*}
and
\begin{equation*}
  \average{\overline{l}^2}
  \equiv
  \int_0^\infty l^2 \bar{p}(l) \diff l.
\end{equation*}
We suppose that $\average{\overline{l}^2} = O(1)$ as $\varepsilon \to 0$.

Expansion of the Laplace transform $\Lapl{\bar{p}}(\varepsilon \lambda)$ in terms of $\varepsilon$ yields
\begin{equation*}
  \Lapl{\bar{p}}(\varepsilon \lambda)
  =
  1
  - \average{\overline{l}} \varepsilon \lambda
  + \frac{1}{2} \average{\overline{l}^2} \varepsilon^2 \lambda^2
  - \ldots{}
\end{equation*}
Hence
\begin{equation*}
  \Lapl{\bar{\phi}}(\varepsilon \lambda)
  =
  \phi_0^{}
  + \phi_1^{} \varepsilon \lambda
  + \phi_2^{} \varepsilon^2 \lambda^2
  + \ldots{}
\end{equation*}
with
\begin{equation*}
  \phi_0^{}
  =
  \frac{1}{\average{\overline{l}}}
  \quad\text{and}\quad
  \phi_1^{}
  =
  \frac{\average{\overline{l}^2}}{2 \average{\overline{l}}^2} - 1
\end{equation*}
(an expression for $\phi_2^{}$ is not needed here).
Therefore, we have formally the asymptotic expansion
\begin{equation*}
  \frac{1}{\varepsilon} \,\bar{\phi} \!\left( \frac{l}{\varepsilon} \right)
  \sim
  \phi_0^{} \delta(l)
  + \phi_1^{} \delta'(l) \varepsilon
  + \phi_2^{} \delta''(l) \varepsilon^2
  + \ldots{}
\end{equation*}
(we consider the expansion in the weak sense).

\section{The Fredholm integral equation}
\label{sec:FredholmEq}

In this section the dependence of values and functions on $\bs{r}$ and $t$ is omitted.

Consider the Fredholm integral equation of the second kind
\begin{equation}
  \label{eq:FredholmEq}
  \left( \Id - \mcal{K} \right) \zeta(\bs{v})
  =
  f(\bs{v}),
\end{equation}
where $\Id$ is the identity operator, and
\begin{equation*}
  \mcal{K} \zeta(\bs{v})
  \defin
  \int_\spaceV K(\bs\varOmega \cdot \bs\varOmega', v, v') \zeta(\bs{v}') \diff\bs{v}'
\end{equation*}
is the scattering operator with the kernel $K(\bs{v}, \bs{v}') \equiv K(\bs\varOmega \cdot \bs\varOmega', v, v')$.
The condition
\begin{equation*}
  \int_\spaceV \int_\spaceV |K(\bs{v}, \bs{v}')|^2 \diff\bs{v}' \diff\bs{v}
  < \infty
\end{equation*}
means that $\mcal{K}$ and its adjoint $\mcal{K}^*$ are Hilbert–Schmidt operators, which implies that they are compact in $L_2^{}(\spaceV)$.
The normalization
\begin{equation}
  \label{eq:KernelNormalization}
  \int_\spaceV K(\bs{v}, \bs{v}') \diff\bs{v}
  = 1
  \quad\Leftrightarrow\quad
  \left( \Id - \mcal{K}^* \right) 1 = 0
\end{equation}
implies that $\zeta^*(\bs{v}) \equiv 1$ is the eigenfunction of the adjoint operator $\mcal{K}^*$ corresponding to the eigenvalue $1$.
Therefore, the Fredholm alternative imposes the solvability condition
\begin{equation}
  \label{eq:SolvabilityCond}
  \int_\spaceV f(\bs{v}) \diff \bs{v}
  = 0.
\end{equation}

The strict positivity of the kernel $K$ and Eq.~\eqref{eq:KernelNormalization} imply that the eigenvalue $1$ is the largest in magnitude and simple eigenvalue of the operator $\mcal{K}$, and the corresponding eigenfunction is positive, \ie, there exist a function $\varPsi(\bs{v}) > 0$ such that
\begin{equation}
  \label{eq:FredhEqHomoSolution}
  \left( \Id - \mcal{K} \right) \varPsi
  = 0,
\end{equation}
see Ref.~\cite{Krasnoselskii:1964}.
Eigenfunctions, corresponding to other eigenvalues, are not nonnegative~\cite{Krasnoselskii:1964}.
Due to isotropic scattering the eigenfunction $\varPsi$ does not depend on the direction $\bs\varOmega$, \ie, $\varPsi(\bs{v}) \equiv \varPsi(v)$.

We need to solve the integral equation with
\begin{equation}
  \label{eq:FreeTermOmega}
  f(\bs{v})
  =
  \bs{a}(v) \cdot \bs\varOmega.
\end{equation}
The solvability condition~\eqref{eq:SolvabilityCond} is satisfied.
To obtain a solution $\zeta$ is expanded in the spherical harmonics $Y_n^m$
\begin{equation*}
  \zeta(\bs{v}) =
  \sum_{n=0}^\infty \sum_{m=-n}^n \zeta_n^m(v) \,Y_n^m(\bs\varOmega)
\end{equation*}
with
\begin{equation*}
  \sum_{m=-1}^1 \zeta_1^m(v) \,Y_1^m(\bs\varOmega)
  \equiv
  \bs\zeta_1^{}(v) \cdot \bs\varOmega
\end{equation*}
(the functions $\zeta_1^m$ are not the components of the vector-function $\bs\zeta_1^{}$).
The kernel is also expanded in the spherical harmonics
\begin{equation*}
  K(\bs\varOmega \cdot \bs\varOmega', v, v')
  =
  \sum_{n=0}^\infty K_n^{}(v, v') \sum_{m=-n}^n  Y_n^m(\bs\varOmega) \,\overline{Y_n^m(\bs\varOmega')}
  \equiv
  \sum_{n=0}^\infty K_n^{}(v, v') \frac{2n+1}{4 \pi} P_n^{}(\bs\varOmega \cdot \bs\varOmega'),
\end{equation*}
where
$P_n^{}$ are the Legendre polynomials.
Due to orthogonality of the spherical harmonics we have
\begin{equation*}
  \mcal{K} \left[ \zeta_n^m(v) \,Y_n^m(\bs\varOmega) \right]
  =
  \left[ \mcal{K}_n^{} \zeta_n^m(v) \right] Y_n^m(\bs\varOmega),
\end{equation*}
where
\begin{equation*}
  \mcal{K}_n^{} g(v)
  =
  \int K_n^{}(v, v') g(v') (v')^2 \diff v',
\end{equation*}
and, therefore,
\begin{equation*}
  \left( \Id - \mcal{K} \right) \left[ \bs\zeta_1^{}(v) \cdot \bs\varOmega \right]
  =
  \left( \Id - \mcal{K}_1^{} \right) \bs\zeta_1^{}(v) \cdot \bs\varOmega.
\end{equation*}
The operator $\left( \Id - \mcal{K}_1^{} \right)^{-1}$ is defined. Indeed, if $\left( \Id - \mcal{K}_1^{} \right) \bs\zeta_1^{}(v) = 0$, then $\bs\zeta_1^{}(v) \cdot \bs\varOmega = \const \,\varPsi(v)$. Therefore, $\bs\zeta_1^{}(v)  = 0$.

As a result, we obtain that the solution of the integral equation~\eqref{eq:FredholmEq} with the free term~\eqref{eq:FreeTermOmega} is
\begin{equation*}
  \zeta(\bs{v})
  =
  c \varPsi(v)
  + \left[ \left( \Id - \mcal{K}_1^{} \right)^{-1} \bs{a}(v) \right] \cdot \bs\varOmega,
\end{equation*}
where $c$ is an arbitrary constant.

\end{document}